 \documentclass[12pt,preprint]{aastex}

\newcommand \hdf {$H\delta_F$}
\newcommand \hb  {$H\beta$}
\newcommand \arcmpt{\hbox to 1pt{}\rlap{\arcmin}.\hbox to 2pt{}} 

\newbox\grsign \setbox\grsign=\hbox{$>$} \newdimen\grdimen \grdimen=\ht\grsign
\newbox\simlessbox \newbox\simgreatbox
\setbox\simgreatbox=\hbox{\raise.5ex\hbox{$>$}\llap
     {\lower.5ex\hbox{$\sim$}}}\ht1=\grdimen\dp1=0pt
\setbox\simlessbox=\hbox{\raise.5ex\hbox{$<$}\llap
     {\lower.5ex\hbox{$\sim$}}}\ht2=\grdimen\dp2=0pt
\def\simgreat{\mathrel{\copy\simgreatbox}}
\def\simless{\mathrel{\copy\simlessbox}}
\newbox\simppropto
\setbox\simppropto=\hbox{\raise.5ex\hbox{$\sim$}\llap
     {\lower.5ex\hbox{$\propto$}}}\ht2=\grdimen\dp2=0pt

\slugcomment{For submission to the Astrophysical Journal Letters}  

\shorttitle{Blue Horizontal Branch Stars in Globular Clusters}
\shortauthors{Schiavon et al.}

\begin{document}


\title{The Identification of Blue Horizontal
Branch Stars in the Integrated Spectra of Globular
Clusters\footnotemark[1]\footnotetext[1]{Based on observations
collected with the CTIO 4m Blanco telescope}} 


\author{Ricardo P. Schiavon}
\affil{Astronomy Department, University of Virginia, P.O. Box 3818, 
Charlottesville, VA 22903-0818}
\email{ripisc@virginia.edu}

\author{James A. Rose}
\affil{Department of Physics and Astronomy, CB 3255, University of North
Carolina, Chapel Hill, NC 27599}
\email{jim@physics.unc.edu}

\author{St\'ephane Courteau}
\affil{Department of Physics, Queen's University, Kingston, ON Canada K7L 3N6}
\email{courteau@physics.queensu.ca}

\author{Lauren A. MacArthur}
\affil{Department of Physics \& Astronomy, University of British Columbia, 
6224 Agricultural Road, Vancouver, BC Canada V6T 1Z1}
\email{lauren@physics.ubc.ca}




\begin{abstract}
A major uncertainty in the spectroscopic dating of extragalactic globular
clusters concerns the degenerate effect that age and horizontal
branch morphology have on the strength of Balmer lines. In this {\it
Letter} we show that the ratio between the equivalent widths of \hdf\
and \hb\ is far more sensitive to horizontal branch morphology than to
age, thus making it possible to break the degeneracy.  We show that it is
possible to distinguish intermediate-age globular clusters from those whose
Balmer lines are strengthened by the presence of blue horizontal branch
stars, {\it purely on the basis of the clusters' integrated spectra}.
The degeneracy between age and horizontal branch morphology can be lifted
with \hb\ and \hdf\ line strengths from spectra with S/N $\simgreat$
30/${\rm\AA}$, which is typical of current studies of integrated
spectroscopy of extragalactic globular clusters.


\end{abstract}




\section{Introduction}

In the last decade, the use of equivalent widths (EWs) of Balmer
absorption lines in the integrated spectra of old stellar populations
(SPs) has become a standard procedure for estimating their light-weighted
mean ages (Worthey 1994). The rationale of the method is to utilize the
sensitivity of Balmer lines to the temperature and luminosity of turn-off
stars, which depend strongly on the age of a SP.  However, a well known
caveat is the existence of old hot stars that may be sufficiently bright
and numerous to boost the EWs of Balmer lines, mimicking the signature
of young stars in the integrated spectra of old SPs. That is the case
of blue horizontal branch (BHB) stars, which are present in such old
systems as Galactic globular clusters (GCs) with a range of metallicities
(see Moehler 2001 for a review), the halo and bulge fields (e.g. Kinman
et al.\ 2000), the cores of Local Group galaxies (Brown et al.\ 1998;
Brown et al.\ 2000; Brown 2003), and old Galactic clusters like NGC~6791
(e.g. Peterson \& Green 1998). So far the only way of unequivocally
distinguishing between an old SP that hosts BHB stars from a truly young
SP is by construction of color-magnitude diagrams (CMDs), which of course
can only be achieved for nearby systems, within the Local Group.

Theoretical studies (Freitas Pacheco \& Barbuy 1995; Lee et al.\ 2000;
Maraston \& Thomas 2000) have shown that the impact of BHB stars on the
integrated spectra of old SPs can be significant, and in particular it may
affect the ages inferred from the EWs of Balmer lines. However, in view
of the lingering theoretical uncertainties regarding the morphology of
the horizontal branch (HB) and its connection with mass loss on the red
giant branch (Catelan 2000), an accurate assessment of the contribution
of BHB stars to the integrated optical light of old SPs still depends
on the availability of CMDs.

In this {\it Letter} we present the first results of an ongoing project
to calibrate our SP synthesis models with high quality integrated spectra
of star clusters. While this has already been attempted by a number of
groups (Beasley et al.\ 2002; Schiavon et al. 2002a,b; Puzia et al.\
2002; Maraston et al.\ 2003; Leonardi \& Rose 2003), our work benefits
from the combination of a homogeneous set of high S/N integrated GC
spectra newly collected by our group, with the HST CMDs of several tens
of Galactic GCs, most of them reaching the GC turn-off, by Piotto et al.\
(2002).  We demonstrate the ability of well-calibrated models, combined
with high S/N spectra, to single out the contribution of BHB stars to the
integrated light of GCs. We use a combination of an Fe-sensitive index
and an index comprised of the ratio of two Balmer line EWs to uniquely
constrain the BHB contribution to the integrated light. This enables us
to distinguish, {\it purely on the basis of their integrated spectra},
truly intermediate-age or young clusters from those which are old,
but whose Balmer lines are strengthened by the contaminating light of
BHB stars. We envisage a direct application of our method to studies
of extragalactic GC systems, where the determination of GC ages and
metal abundances can yield insight into the star formation and merger
histories of the host galaxies (e.g. Cohen, Blakeslee \& C\^ot\'e 2003;
Larsen et al. 2003; Hempel et al.\ 2003).


\section{Observations} \label{obs}

We collected integrated spectra for 40 Galactic GCs with the
R-C spectrograph on the 4m Blanco telescope at CTIO in April 2003.
We scanned a 5\arcmpt5-long slit across the core diameter of each target
GC. For the GCs located towards the bulge, on-target exposures were
interspersed with raster scans of adjacent sky regions $\sim$5\arcmin\
away from the GC centers. The instrumental setup consisted of grating
KPGL1, with 632 l/mm, and a Loral CCD with 3k $\times$ 1k 15$\mu$m-sized
pixels. The resulting spectra cover the range 3365--6435${\rm\AA}$
with 2.8${\rm\AA}$ FWHM resolution. The exposure times were scaled
to yield spectra with S/N $\simgreat$ 100 per resolution element. Data
reduction used standard IRAF routines for longslit spectra. Final
1D integrated spectra were obtained by extracting an aperture covering
the core diameter along the slit direction. Since the exposures were
trailed over a core diameter perpendicular to the slit, the resulting
1D spectra are representative of the stellar content in the cores of the
GCs. The spectra were flux calibrated using observations of flux
standards taken throughout the observing run. The 1D integrated spectra
had their resolution degraded to match that of the Lick/IDS system,
and EWs of absorption lines were measured following the definitions
of Worthey et al.\ (1994) and Worthey \& Ottaviani (1997). Consistency
with the Lick/IDS index system was achieved by comparing EWs measured in
spectra of standard Lick/IDS stars, taken throughout the observing run,
with standard values from Worthey et al.\ (1994).


\section{Effect of Blue Horizontal Branch Stars on Balmer Lines}
\label{res}

\begin{figure}
\plotone{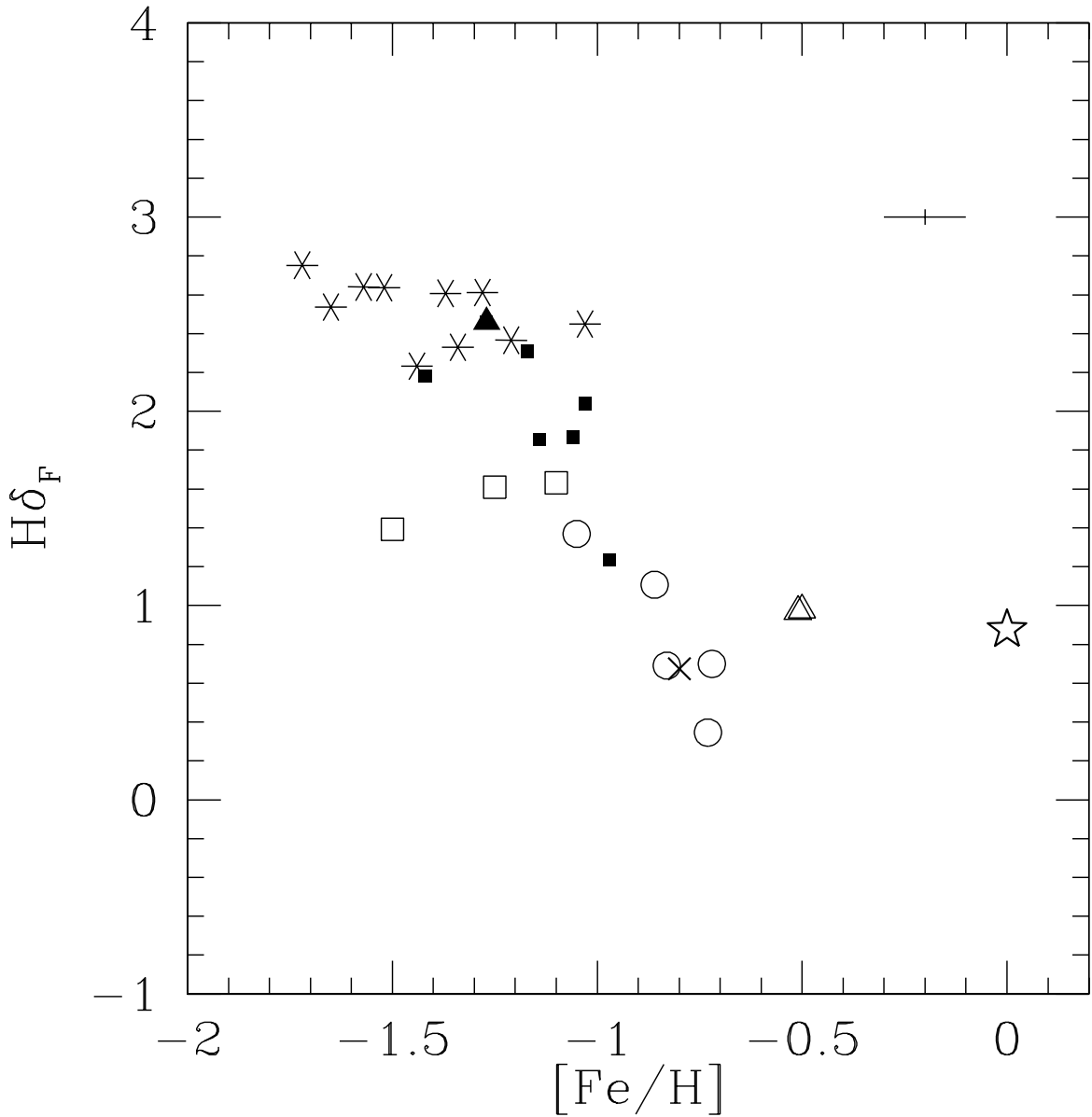}
\caption{The sample of GCs analysed in this paper: GCs with mostly blue
HBs are depicted as asterisks, open circles represent those with {\it
strictly red} HBs, GCs with {\it mostly red} HBs are labelled as open
squares, and GCs with intermediate HB morphologies are represented by
small filled squares.  The two open triangles stand for NGC~6388 and
NGC~6441, the filled triangle represents NGC~5904, and the $\times$
stands for 47 Tuc. The star represents M67. The effect of HB morphology
on $H\delta$ can be seen as the $\sim$1 ${\rm\AA}$ vertical spread for
--1.5 $<$ [Fe/H] $<$ --0.9. Here and in all subsequent figures, typical
error bars are displayed in the upper right corner.
} \label{fig1} 
\end{figure}

In order to highlight the impact of HB morphology on the integrated light,
we restrict our analysis to GCs with [Fe/H] $\simless -0.5$, because at
higher metallicities all GCs have similar red HBs, therefore not adding
much to our discussion. A plot of \hdf\ vs. [Fe/H] for this sub-sample is
shown in Figure \ref{fig1}. For most of the GCs, [Fe/H] values were taken
from Kraft \& Ivans (2003).  Those values were supplemented, whenever
needed, by data from Carretta \& Gratton (1997), Idiart, Th\'evenin \&
Freitas Pacheco (1997), and Harris (1996). Throughout this {\it Letter}
GCs are subdivided in terms of HB morphology, with $x_{HB}={B-R\over
B+V+R}$, as follows: GCs with {\it mostly blue} HBs ($x_{HB} > +0.4$),
GCs with {\it mostly red} HBs (--$0.9 < x_{HB} < -0.4$), GCs with {\it
strictly red} HBs ($x_{HB}$ $<$ --0.9), and GCs with intermediate HB
morphologies (--$0.4 < x_{HB} <$ +0.4). Values of $x_{HB}$ were taken
from Borkova \& Marsakov (2000). In all the figures, a few landmark
clusters are highlighted with special symbols (see caption of Figure
\ref{fig1}). Those are 47 Tuc, a metal-rich GC with [Fe/H] $\sim$ --0.8
and a red HB; M67, a $\sim3.5$ Gyr old, solar metallicity, open cluster
(data from Schiavon, Caldwell \& Rose 2004); NGC~6388 and NGC~6441, the
highest metallicity ([Fe/H] $\sim$ --0.6) Galactic GCs known to host BHB
stars (Rich et al.\ 1997), and NGC~5904, a metal poor GC with [Fe/H]
$\sim$ --1.27 (Kraft \& Ivans 2003) and whose HB has a sizeable blue
extension (Piotto et al.\ 2002).  Our sub-sample spans a wide range in
metallicity and HB morphology; most notably, GCs with --1.5 $\simless$
[Fe/H] $\simless$ --0.9 are strongly affected by HB morphology, in the
sense that \hdf\ in strictly blue-HB GCs is stronger by $\sim$1 ${\rm\AA}$
than in those with strictly red HBs at the same [Fe/H].

\begin{figure*}
\plottwo{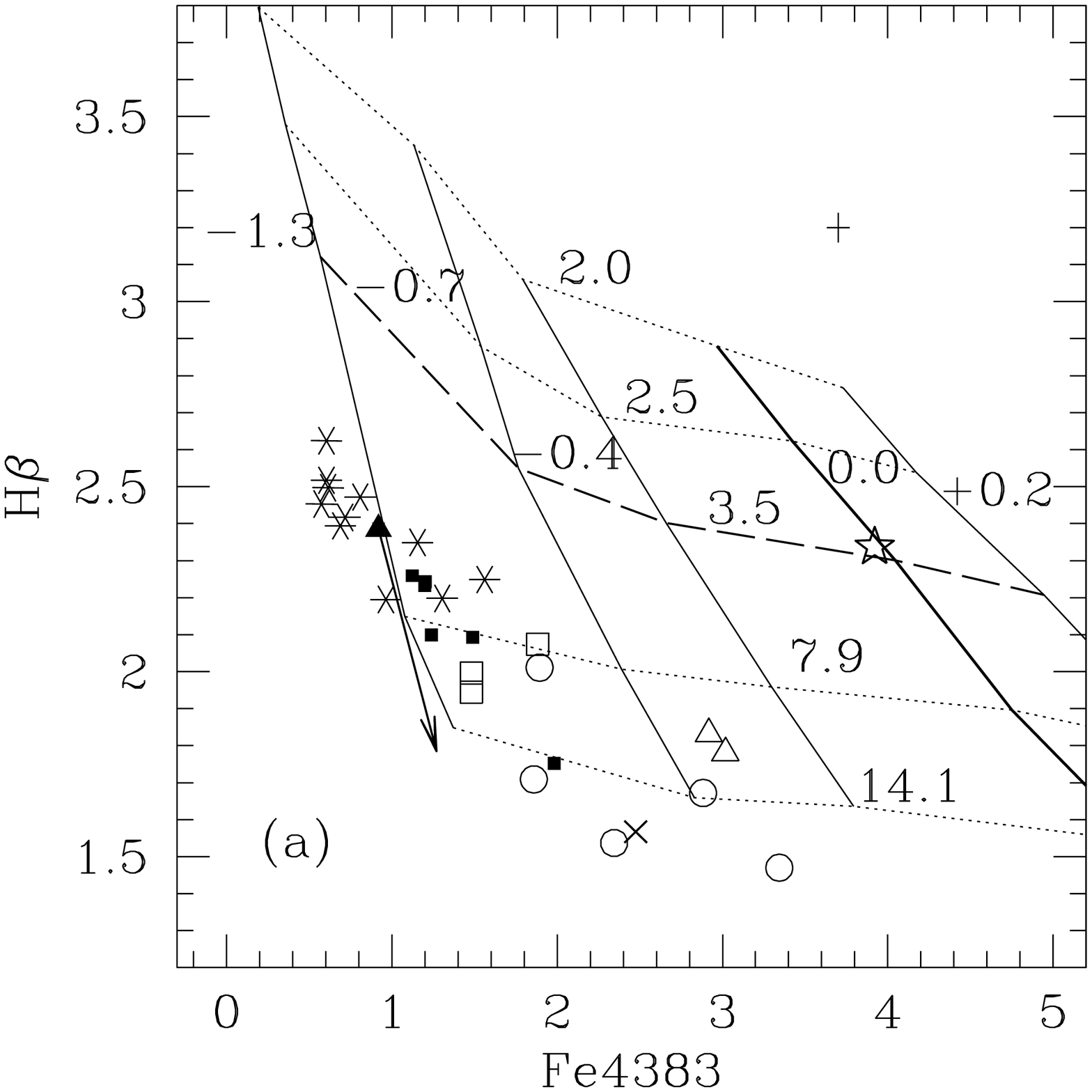}{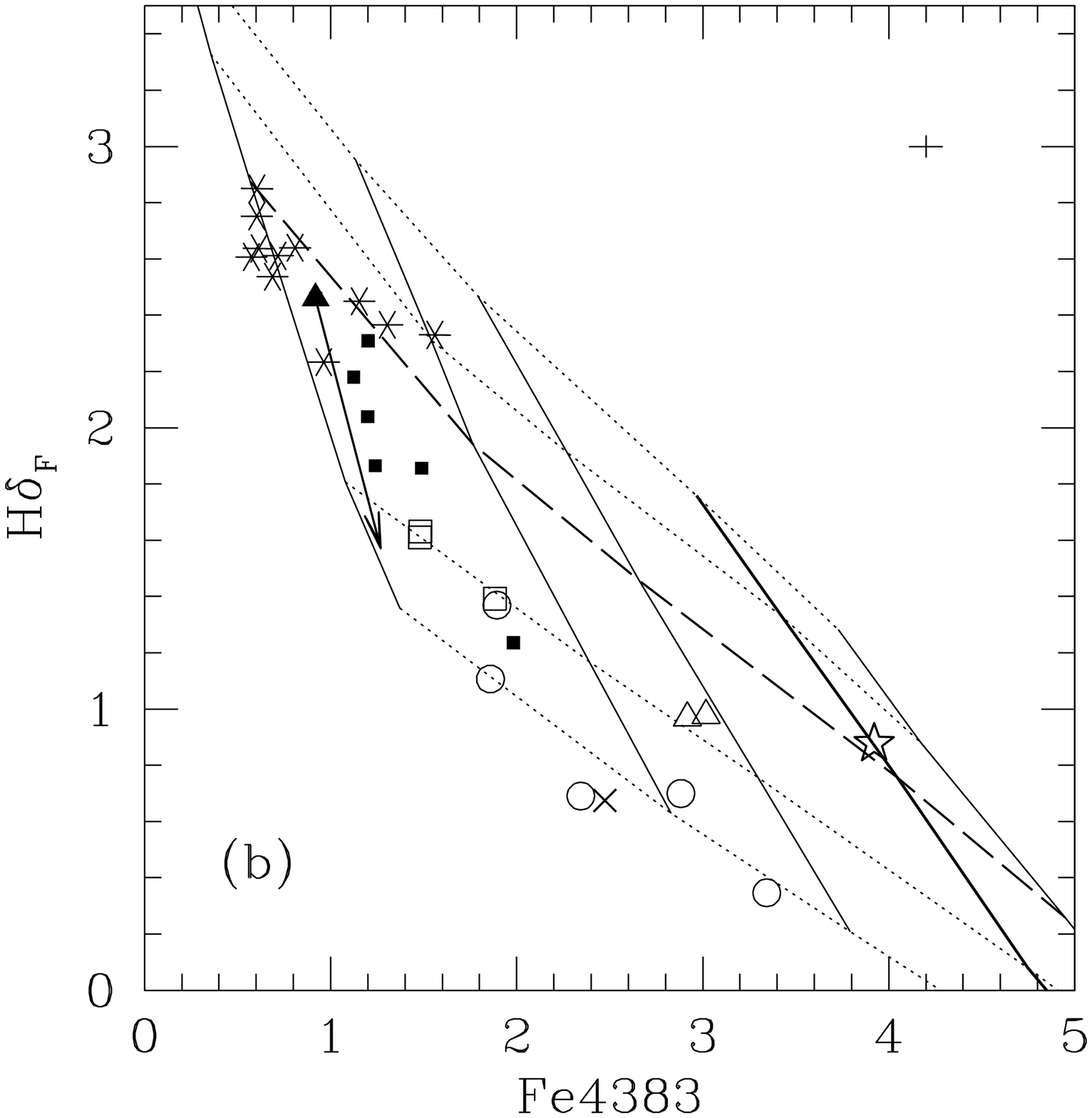}
\caption{Comparison between data and models for single SPs in Balmer
vs. Fe4383 plots. Symbols are the same as in Figure \ref{fig1}. Dotted
lines connect same-age models, while isometallicity lines are solid. Model
lines are labelled with age and [Fe/H] values in panel (a). As a guide to
the eye, the nearly vertical line for [Fe/H]=0 (solid), and the nearly
horizontal line for 3.5 Gyr (dashed) are both thicker. In both panels,
GCs with blue HBs tend to appear younger than those with red HBs, even
though they are all old. GCs with blue HBs also look younger according to
$H\delta_F$ (panel b) than according to $H\beta$. The arrows indicate how
the indices for NGC~5904 change when BHB stars are artificially removed.
} \label{fig2} \end{figure*}

In Figure \ref{fig2} we contrast the data with our model predictions for
Fe4383 and Balmer lines in single SPs. The Balmer indices were chosen to
provide a large baseline and because model predictions were available for
them at the time of this analysis. The models used are those by Schiavon
(2004, in preparation, but see Schiavon et al.\ 2004). 
For our present purposes, derivation of absolute ages
and [Fe/H]s is not essential, and we focus our discussion on relative
spectroscopic ages.

The main issue we address in this {\it Letter} is evident in Figure
\ref{fig2}: GCs with mostly blue HBs (asterisks) have stronger Balmer
lines and thus appear younger than GCs with red HBs (open circles) even
though they are all equally old (e.g. Rosenberg et al.\ 2002). This
effect has been pointed out by many (e.g. Freitas Pacheco \& Barbuy
1995; Lee et al.\ 2000; Beasley et al. 2002; Maraston et al.\ 2003)
and results from the fact that our SPS models do not account for BHB
stars, which have strong Balmer lines and mimick a younger age. This
is a source of confusion for cluster age estimation based on the EWs of
Balmer lines.  Furthermore, while the ages of GCs with red HBs do not vary
according to different Balmer lines, the same is not true for GCs with
blue HBs: their mean ages are $\sim$3.5 and $\sim$6 Gyr as given by \hdf\
and \hb, respectively. The same effect is also observed in the data for
NGC~6388 and NGC~6441 (open triangles), whose inferred ages are $\sim$8
or $\sim$11 Gyr according to \hdf\ or \hb, respectively. In contrast,
for M67, which is a cluster of truly intermediate age ($\sim$3.5 Gyr;
Schiavon et al.\ 2004) and whose spectrum is free from the influence
of BHB stars, the inferred age is independent of the age indicator. We
believe that these effects are caused by the increasing contribution
of BHB stars to the integrated GC light at bluer wavelengths, such that
\hdf\ is more affected than \hb. As a result, the GC spectroscopic ages
look younger according to \hdf\ than \hb.


To verify this hypothesis we computed integrated indices for NGC~5904
directly from its CMD in two different ways: first by taking all the stars
into account, and then removing all the BHB stars in order to assess
their effect on absorption line EWs. This is the most straightforward
and model independent way of assessing the contribution of BHB stars to
the integrated GC light. We chose NGC~5904 because its HB is mostly blue
and its CMD is indicative of an old age. The colors and magnitudes of
NGC~5904 stars were transformed into $T_{\rm eff}$ and $\log g$ using the
calibrations described in Schiavon et al.\ (2002a) and assuming [Fe/H]
= --1.27 (Kraft \& Ivans 2003). The resulting stellar parameters were
used as input to our fitting functions in order to predict line indices
for each star in the cluster CMD. The latter are integrated throughout
the CMD, with weights based on the photometry and stellar counts. The BV
photometry, its completeness as a function of V magnitude, the distance
modulus [$(m-M)_V = 14.46$], and the reddening towards NGC~5904 [$E(B-V)
= 0.03$] were taken from Piotto et al.\ (2002). The result is displayed
by arrows in Figure \ref{fig2} that indicate how index values change
when BHB stars are artificially removed from the CMD. The position of
the cluster in both plots is shifted towards an old age ($\sim$11--14
Gyr), which corroborates our working hypothesis. Furthermore the above
correction does not alter the agreement of our model predictions with
[Fe/H] determinations from spectroscopic analysis of individual stars
from NGC~5904 ([Fe/H] $\sim$ --1.3; Kraft \& Ivans 2003).

In summary, we suggest that the different spectroscopic ages inferred
from $H\beta$ and $H\delta_F$ are due to the stronger impact of BHB
stars on the latter. Below, we use this observation to disentangle the
degeneracy between the effects of HB morphology and age on Balmer lines
measured in the integrated spectra of GCs.



\section{Detection of Blue Horizontal Branch Stars in Globular Clusters
} \label{detect}

\begin{figure}
\plotone{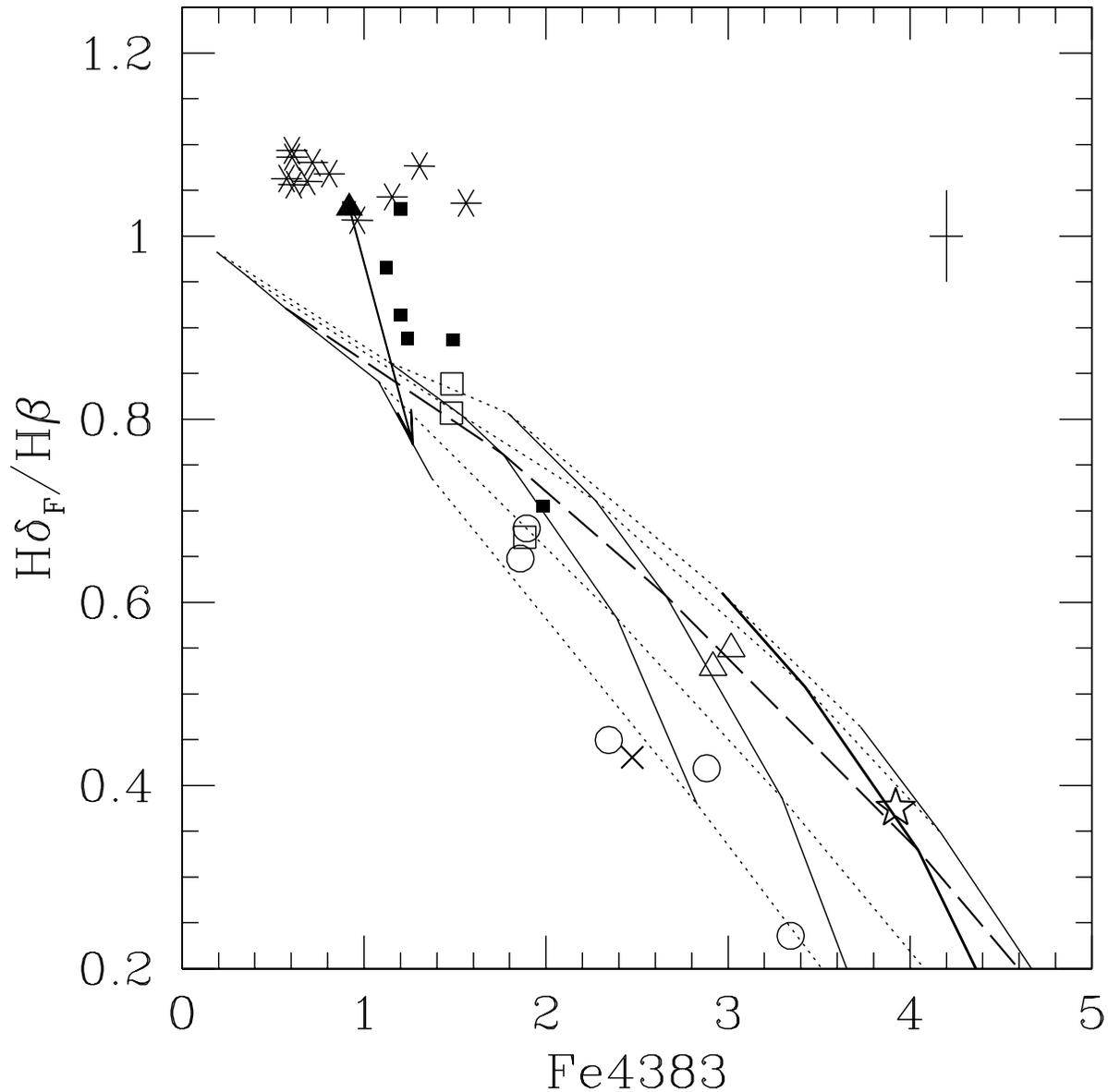}
\caption{The data are confronted with the models in an Fe vs. Balmer
line ratio plane. Symbols, lines and arrow have the same meaning as
in the previous figures. GCs with blue HBs are greatly displaced from
the model locus, indicating ages which are much younger than those
found in Figure \ref{fig2}. In constrast the ages and [Fe/H]s for M67
and GCs with strictly red HBs are consistent with those obtained from
Figure \ref{fig2}. The ratio $H\delta_F/H\beta$ is more sensitive to HB
morphology than to age, and therefore helps in breaking the degeneracy
between the two parameters.
} \label{fig4}
\end{figure}

We have identified a distinctive signature of BHB stars in the integrated
spectra of Galactic GCs. The obvious implication is that we can devise
an indicator that identifies their presence in a given GC, solely on the
basis of absorption line EWs. Such an indicator would be most useful for
work on distant GCs, for which CMDs cannot be obtained, by providing the
means to distinguish truly intermediate-age GCs from old GCs with BHB stars.
The best indicator we found is the ratio between \hdf\ and \hb.  Figure
\ref{fig4} displays the main result of this {\it Letter}, where we plot
the data in the $H\delta_F/H\beta$ vs. Fe4383 plane, overlaid with our
models. It can be seen that GCs
with mostly blue HBs are displaced far above the locus occupied by the models,
indicating ages substantially younger than $\sim$2 Gyr, in stark constrast
with the ages inferred in Figure \ref{fig2}. In fact $H\delta_F/H\beta$
is more sensitive to HB morphology than age, given that the separation
in $H\delta_F/H\beta$ between GCs with red and blue HBs is about
twice that between models for 2 and 14 Gyr single SPs.

GCs with strictly red HBs, such as 47 Tuc, are consistent with the ages
and metallicities inferred from Figure \ref{fig2}. As expected, GCs
with mostly red or intermediate HBs lie somewhere in between the loci
occupied by GCs with extreme HB morphologies.  Note that the data for
M67, which has a truly intermediate age, are (as in Figure \ref{fig2})
consistent with its known age and [Fe/H]. We conclude that the ratio
$H\delta_F/H\beta$ discriminates between younger clusters and those
whose Balmer lines are strengthened by the presence of BHB stars.

The high S/N of our spectra make it relatively easy to detect the
influence of the BHB population on the $H\delta_F/H\beta$ index. For
instance, for NGC~5904 which has $H\delta_F/H\beta = 1.04 \pm 0.05$
(1 $\sigma$), the value expected to be consistent with the age estimated
from \hb\ from Figure \ref{fig2} ($\sim$ 6 Gyr) would be $H\delta_F/H\beta
\sim 0.84$, or $\sim$4 $\sigma$ lower than the measured value. A useful
quantity for observers is the minimum S/N needed to distinguish between
a GC with BHB stars and a young GC. This was estimated by computing index
errors as a function S/N, assuming pure Poisson noise. In the case above,
one needs $H\delta_F/H\beta$ measured to better than 0.2 which, for a
2-$\sigma$ detection requires a S/N of roughly 30--40 per ${\rm\AA}$ at
$\simeq$ 4100 ${\rm\AA}$, a value that can be achieved with reasonable
integrations of extragalactic GCs with 8-10m class telescopes (e.g. Cohen
et al.\ 2003; Larsen et al.\ 2003; Hempel et al.\ 2003).

\section{Final Remarks} \label{final}

In this {\it Letter} we propose a method to detect the signature of
BHB stars in the integrated spectra of GCs. If \hdf, \hb, and an Fe line
are measured accurately, one can infer the presence of BHB stars in a
given GC if the age indicated by $H\delta_F/H\beta$ is substantially
younger than that indicated by $H\beta$. For intermediate-age clusters,
the ages indicated by the different indices are the same
within the uncertainties. For this method to work, {\it 
SP synthesis models employed must predict correct
ages and metal abundances for systems with truly intermediate/young
ages}. We showed that our models predict the correct age and [Fe/H] for
M67. While the case for applying this method to disentangle age from HB
morphology seems strong for single-age systems, {\it it is not clear that
it should succeed when applied to galaxies},
where multiple turnoffs due to a complex star formation history cannot
in principle be distinguished from the combination of an old turnoff
with other kinds of warm stars like BHB stars.


\acknowledgments

We thank Francesca de Angeli and Manuela Zoccali for providing us with
data on the completeness of the photometry for NGC~5904. Ruth Peterson,
Duncan Forbes, Robert Proctor, and an anonymous referee are thanked
for useful suggestions. R.P.S. acknowledges financial support from
NSF grant AST 00-71198 to the University of California, Santa Cruz,
and from HST Treasury Program Grant GO-09455.05-A to the University
of Virginia.  S.C. and L.A.M. acknowledge support from the National
Science and Engineering Research Council of Canada, and J.A.R. from NSF
grant AST-99000720 to the University of North Carolina.


\clearpage










\end{document}